\documentclass[10pt,titlepage]{article}

\newcommand{\Section}[1]{\vspace{-8pt}\section{\hskip -1em.~~#1}\vspace{-3pt}}

\begin{document}

\title{The Causet Mechanism for the Creation
of Energy\thanks{This paper is dedicated to Rafael Sorkin and will appear
in the volume published to honor him on his $60^{th}$ birthday.}}

\author{R. Brout \\ Facult\'e des Sciences, Universit\'e Libre de Bruxelles and
\\ The Perimeter Institute for Theoretical Physics, Waterloo,
Canada}
\date{}

\maketitle

\Section{Introduction}

Perhaps the most urgent and most vexing problem that confronts
modern physics is the elucidation of the nature of space-time,
otherwise put, the reconciliation of quantum mechanics with gravity.
How does the expected quantum graininess at the planckian scale give
way to the successful continuum theory of general relativity at much
larger scales?

At present, systematic efforts, notably string theory and/or loop quantum
gravity, remain incomplete in this quest. Therefore, for the nonce, we
must rely on our ``physical intuition" acquired from experience in dealing
with situations more amenable to our human limitations. In this, we resort
to phenomenology, a priori hypotheses designed to handle problems which,
by their nature, force us into physics at the planckian scale. The most
dramatic of these is encountered in the backward extrapolation of
cosmology towards inflation in the quest for a causal universe and thence,
presumably, to creation itself. Less dramatic, but also fascinating, is
the nature of black hole evaporation, once more concerned with the
creation of matter accompanied by an increase of volume of observable
space, in this case the Schwarschild space exterior to the black hole
horizon which is occasioned by the reduction of the black hole mass.

The present contribution, in honor of Rafael Sorkin, is based on Sorkin's
causet phenomenology. Section 2 contains a brief review of the causet
mechanism for the creation of dark energy, Ref.\cite{1}. Then in Section
3, it is argued that causets could apply to inflation as well as to dark
energy in the adiabatic era, Ref.\cite{1}, whereupon one is led to
speculate that this latter is the fluctuating remnant of the vacuum energy
responsible for inflation. The last section indicates how causets may be
applicable to black hole evaporation.

We close this introduction with a question. When space expands, as
in cosmology, generally speaking there are two options. Is there an
underlying metric manifold which metric encodes the expansion? Or
does one create space as in the causet phenomenology? The latter
motivates loop quantum gravity; the former is the more traditional
point of view of general relativity. We shall here follow the
second alternative. In this, it is well to point out that even if
one pursues the more traditional point of view, one still cannot
avoid something of the causet idea, related to the so called
space-time foam, necessitated by the need for a reservoir of modes
of quantum fields since elsewise the density of modes would decrease
during the expansion. So whichever way, one is forced to speculate
in terms of a priori hypotheses.

\begin{verse}
Let imagination roam fancy free \\ To speculate what the world might
be.
\end{verse}

\Section{The Causet Scenario of Dark Energy in the Adiabatic Era}

It is postulated that as the universe expands, space is delivered as
discrete lumps of planckian dimensions in space-time. These are
distributed at random with a mean density which is planckian in
space-time. The word causet is used because only the causet elements
situated within the backward light cone of a given event can influence
happenings around that event. The happenings of interest are those which
cause variations of energy near the event (near and around are presumed to
mean within a planckian space-time interval). To this end, one postulates
that the causet elements are the sites wherein vacuum energy is given to
or taken from those degrees of freedom within the universe whose energy
causes the expansion, as dictated by the condition of energy constraint.
This energy is created or annihilated as space is created. One may suppose
that there are ``hidden" degrees of freedom within the causet element
whose energy content does not affect the expansion. In Section 3 we shall
draw upon an analogy in matter physics which may help one to visualize
this process. For the moment, following Ref.\cite{1}, we sketch how this
scenario can deliver the requisite dark energy density. In particular, it
explains why the energy density of vacuum ($\rho_\Lambda =$ dark energy
density) is the same order of magnitude as the energy density of matter
($=\rho_M$) during the adiabatic era.

We now present the beautiful argument of Ref.1. One introduces the energy
density ($\rho_\Lambda(t)$) coupled to the macroscopic expansion through
the energy constraint: $H^2(t)=\rho_\Lambda(t) + \rho_M(t)
=\rho_\mathrm{Tot}(t)$. We have set $m_\mathrm{pl}=1$ and all constants of
$O(1)$ are set equal to $1$. $H(t)$ is the Hubble constant at time $t$. We
repeat: any energy associated to possible degrees of freedom within the
causet is not contained in $\rho_\Lambda(t)$. Rather, consider
$\rho_\Lambda$ to be a conventional type of zero point field energy
(including gravitational and other interactions) which is born or
annihilated through interactions between conventional cisplanckian
configurations of fields and degrees of freedom in the causet elements.

The space-time volume of the backward light cone of an observer at proper
time t, during the adiabatic era, is easily shown to be $O(H^{-4}(t))$.
For that observer, one must compute the effective vacuum energy that is
due to the causets within that zone. Since the causet density is $O(1)$
the number of such elements $(= N(t))$ is $O(H^{-4}(t))$. Each element is
the site of  a random exchange with vacuum; hence the net accumulated
vacuum energy density, in the sense of the preceding paragraph, is $\pm
O(\sqrt{N(t)}/ H^{-4}(t)) = O(H^2(t))$. Since $H^2 = \rho_\mathrm{Tot}$,
one sees $\rho_\Lambda(t) = O(\rho_\mathrm{Tot})$ and it is presumed that
$\rho_\mathrm{Tot}$ has a significant portion which is $\rho_M(t)$ in the
adiabatic area.

This scenario has been put into more quantitative form as a stochastic
equation which gives $\rho_\Lambda(t_{i+1})$ in the $(i + 1)^\mathrm{th}$
time slice in terms of $\rho_\Lambda(t_i)$ in the $i^\mathrm{th}$ slice of
the backward light cone. Some runs duplicate phenomenology remarkably
well, Ref.\cite{1}.

In this, only the energy constraint is used. Since $\rho_\Lambda$
varies with $t$ (in fact in both space and time, but space is taken
to be sufficiently homogenous to ignore spatial variation), one
should include the acceleration equation of cosmology as well, along
with an equation of state. To assimilate $\rho_\Lambda$ to a
cosmological constant is an approximation that will require
justification in future work. In what follows, we assume its
legitimacy.

An essential assumption in Ref.1, which is truly profound, is the
characterization of the ``target" value of $\rho_\Lambda$ in the
adiabatic era, to wit: as $\rho_M\to 0$, it is assumed that
$\rho_\Lambda\to 0$. Thus the asymptotic universe is empty and
quiescent ($H=0$).

But quantum mechanics cannot tolerate such tranquility. Fields fluctuate
and gravity has an unbounded spectrum from below as is manifested in the
energy constraint where $-H^2$ is an average negative energy density
associated with gravity when applied to the cosmological expansion. Thus,
this empty quiescent universe should be considered a metastable state. A
fluctuation from it can seed a new universe. An example of such a seeding
is in  Ref.2, Ch.8. Another is the concept of chaotic inflation conceived
in Ref.3 and later in Ref.4. In fact the very idea of inflation and a
causally created universe was so conceived, Ref.\cite{5}.

Therefore, the causet phenomenology of dark energy, taken together
with the assumption that $\rho_\Lambda\to 0$ as $\rho_M\to 0$ on the
average, forces one to speculate on the physics of vacuum, creation
and inflation. These speculations are presented in Section 3 where
we shall argue that the causet mechanism of dark energy creation in
the adiabatic era is the fluctuating remnant of the causet energy
creation process of the inflationary era.

\Section{Inflation and Creation}

It would be nice if one could extend the causet-mode exchange idea
to the earliest times. The inflationary era was introduced into
physics to be able to conceive the universe as a causal response to
a seed which is a vacuum fluctuation. If we retain this view of
inflation, and in the following paragraphs this shall be our point
of view, the causet scenario of the adiabatic era must be modified;
for if the seed arises from vacuum there is no possibility for the
causet element to remove energy, but rather only to give energy to
the nascent universe. A further consideration which prompts us to
modify the causet scenario when applied to inflation is the vast
difference in magnitude of $\rho_\Lambda$ between the adiabatic
epoch where
$\rho_\Lambda^\mathrm{Now}=O(\rho_M^\mathrm{Now})<O(10^{-100})$ and
inflation where $\rho_\Lambda = O(10^{-10}$). The ratio of Hubble
constants for example is $H^\mathrm{Now} / H^\mathrm{inflation} = O
(10^{-50})$.

To come to grips with this question it is useful to inquire into the
possibility of modeling a mechanism for causet-vacuum energy exchange. To
this end there is a useful analogy which may serve as a guide, the tight
binding mechanism for the generation of electron bands in metals from
atoms.

An isolated atom has all its electrons localized in the vicinity of
its nucleus. But when the atoms approach each other the wave
functions of the outer (valence) electrons overlap and the wave
functions develop into delocalized bands which, by translational
symmetry, are classified by momenta rather than localized orbitals.
The inner or core electrons, in very good approximation, remain
localized. Thus we have a simple model in which degrees of freedom
of the same type of stuff are classified into two widely different
types of configurations. Let us now return to field theory.

In usual quantum field theory, the fields are developed in terms of
modes characterized by their momenta. This approach ceases to be
valid when the momentum becomes planckian for then the field
coupling to gravity induces strong coupling among the modes and the
momentum of a planckian mode is a concept that loses its usefulness.
It makes no more sense to describe configurations of fields at the
planckian scale in terms of modes and their momenta than it does to
describe a liquid in terms of the momenta of its constituent atoms.
Modes at the planckian scale are strongly coupled as are the
molecules of a liquid and one must seek an alternative way to
describe field configurations at that scale. Degrees of freedom of
fields at the planckian scale could well fold up into localized
structures. These could be the causet elements. They can also be
called elements of space-time foam, sometimes envisaged as black
hole in character.

In a quantum theory of gravity where space-time can be conceived as
field variables as well, these causet elements of planckian
dimensions then could correspond to Sorkin's conception of how space
is created to describe the cosmological expansion.

Long wavelength modes still exist since their mutual interactions
are weak. So the same kind of stuff, at the small length scale, can
be causets, and at long length scale, are describable in terms of
the conventional modes of field theory on a background.

This picture accommodates well to the problem posed by the dilution of
modes induced by the cosmological expansion. The problem, as usually
stated, is that modes must be cut off at planckian momenta, because they
are then strongly coupled.  But the cut-off decreases as the inverse of
the scale factor. So they must be replenished from a reservoir. In the
causet scenario this is automatic. The causets remain at constant density
and the modes cut off at the length scale that separates the causet
elements. No dilution! The visualization offered by the tight binding
mechanism of bands helps in this conception. One just makes larger and
larger crystals, more and more atoms, as well as more and more electrons
in the modes (bands). Their number is (Volume) $\times$ (cut-off
momentum)${}^3$, Ref.\cite{3}, where the cut-off momentum stays fixed during the
expansion.

This image also could give some notion of how it is that the degrees of
freedom sequestered in a causet element have no influence on the
macroscopic dynamics such as the Hubble constant that figures in the
energy constraint. In the tight binding analogy, the pressure exerted by
the electron gas  in many metals comes from the electrons in the bands.
The core electrons have no role. (For metals where the cores are far one
from the other, such as the alkalis, the bulk modulus is well approximated
by that of the free electron gas formed from the valence electrons). So it
is that field degrees of freedom within the causet elements can have no
impact at the macroscopic level. It is only through the microscopic
processes of the energy exchange with the mode system that indirectly they
have influence. That is the substance of the causet explanation of dark
energy, an energy that is transferred to the mode system in vacuum,
exchanged in and out of the causet elements.

Unfortunately, these concepts are not sufficient to rationalize the
existence of  a quiescent metastable vacuum. But since in the following
paragraphs we shall build a scenario of cosmogenesis and subsequent
inflation from such an initial state, it behooves us to make some remarks
in its defense.

There is some element of self consistency in the state $\rho_M=0$,
 $\rho_\Lambda=0$, hence $H=0$. No net temporal variation of the
average metric is consistent with no net creation of quanta, hence $\rho_M
= 0$ thence, $\rho_\Lambda = 0$ since $H=0$. In the causet phenomenology
we envision causet elements whose population fluctuates leaving no net
expansion of space and no net exchange of energy to the cisplanckian
world. Perhaps this occurs in a mathematical framework which at present,
at least, is beyond our ken, such as the elements of spin foam or the
$SU_2$ matrices of Kodoma or the high energy configurations of strings
which are black hole in character or, or, $\dots$. Whatever these
fluctuating bits of space-time may be, when taken together with the
ephemeral configurations of matter fields which they carry, in this
metastable state they transfer no energy to our universe in the
macroscopic sense.

To seed a universe we must appeal to a collective type of phenomenon
wherein a number of causet elements act coherently to give a macroscopic
sense of expansion, hence a gravitational average energy due to the
expansion which is $-H^2$.

Then the causet can start to deliver energy to the cisplanckian sector in
that part of space where a sufficient number have agglomerated to make a
seed. Modes form and at the same time pick up vacuum energy. At this early
stage we expect $\rho = O(1)$ since the causets deliver their energy as
well as space to the seed. The seed causes space to expand rapidly with
$H=O(1)$ and a macroscopic universe composed of causets and the modes
which develop from them comes into being. One may imagine that this rapid
expansion does not allow time enough for a significant amount of vacuum
energy to be reabsorbed onto the causets as they do in the quasi
stationary situation of the adiabatic era, i.e., $\rho_\Lambda = O(N/N) =
O(1)$ in the early universe rather than
$\rho_\Lambda=O(\frac{1}{\sqrt{N}})$ as in the adiabatic era.

Our present knowledge of the early universe is well accounted for by
the inflationary scenario. And the implementation of the inflation
through the inflaton scenario has been highly successful in
accounting quantitatively for the CMBR fluctuations. Furthermore the
inflaton has the advantage that it has its own demise built into the
formulation of the concept and its equation of motion. Is it
possible to rationalize the existence of an inflaton from the seed?
And is it possible to offer some explanation for its mass, $\mu
(=10^{-5})$ in terms of the causet concept? To answer these
questions is extremely difficult and these next paragraphs are
speculative in the extreme.

We postulate that the inflaton comes into existence as a collective mode
of fluctuation once the initial seed has become well organized so as to
justify the use of the macroscopic notions of cosmology, a homogeneous
space as a viable zero'th order approximation which expands through a
Hubble constant that depends on time only throughout the patch that
contains this small universe. This macroscopic system is now endowed with
modes and causet elements which interact. Having emerged from a planckian
seed with $H=O(1)$, it will take a while for that initial expansion to
slow down and then turn over to the adiabatic stage. This period between
the formation of the seed and the adiabatic expansion is the inflationary
epoch. Phenomenology indicates that its expansion rate is slower than
planckian, say $H=O(10^{-5})$ and the inflaton mass follows suit
$\mu=O(10^{-5})$ if the mean value of the inflaton amplitude, $\phi$, is
taken to be $O(1)$.

We postulate that the inflaton comes into existence as a collective mode
of fluctuation once the initial seed has become well organized to justify
the use of the macroscopic notions of cosmology, a homogeneous space as a
viable zeroth order  approximation which expands through a Hubble
constant, that depends on time only, throughout the patch that contains
this small universe.  This macroscopic system is now endowed with modes
and causet elements which interact.  Having emerged from a planckian seed,
with $H=O(1)$, it will take a while for that initial expansion to slow
down and then turn over to the adiabatic stage.  This period between the
formation of the seed and the adiabatic expansion is the inflationary
epoch.  Phenomenology indicates that its expansion rate is slower than
planckian, say $H=O(10^{-5})$, and the inflaton mass follows suit, $\mu=
O(10^{-5})$, if the mean value of the inflaton amplitude, $\phi$, is
taken to be $O(1)$. That $\phi=O(1)$ is a natural initial condition is
indicated by the mechanism initiated by the seed wherein the only scale is
O(1).

It is less simple to rationalize $\mu=O(10^{-5})$. We sketch below  a
speculation given in Ref.\cite{broutx}, now taken over for causets.

Within the inflating patch there is the planckian graininess at the
smallest scales.  It is proposed that there exists a local equilibrium
between the degrees of freedom sequestered within the causets and the
modes just as is there is during the adiabatic epoch and ultimately in the
metastable vacuum.  It is maintained by the cis-trans exchanges to and
from the causets, but now in the background $\phi = $ const. $=O(1)$
during the greater part of the inflationary epoch.  This is taken to be a
rapid process whose characteristics are in good approximation background
independent.  Rapid means on the scale of the inflaton decay from $\phi
= O(1)$ towards the reheating era and the subsequent adiabatic expansion.

It was proposed in Ref.\cite{broutx} that the fluctuations about this
effective two fluid quasi-equilibrium were those of a massy acoustic
degree of freedom, massy because neither the density of modes nor of
causet elements is conserved.  It was argued that $\mu^2=M^2p$ where $M$
is the mass scale of the matrix element for the cis-trans exchange and $p$
the dimensionless probability encoded in that matrix element.  One may try
to estimate $p$ by making use of the parameter $\alpha$ introduced in  the
stochastic process of Ref. 1.  There it was found that $\alpha$ was
tightly constrained to be $O(10^{-2})$. With $p= O(10^{-2})$ and
$\mu=O(10^{-5})$ one finds $M =O( 10^{-4})$. It is difficult to interpret
these parameters save that one might expect $M=O(1)$. In
Ref.\cite{broutx}, black hole processes taking place in vacuum were
speculated upon as a possible interpretation.

In summary, the creation -inflation scenario proposed from the causet
phenomenology is: ~1)  formation   of an initial seed from causet elements
which function to create both space and vacuum energy with $\rho_\Lambda
=O(1)$; ~2) the subsequent inflation driven by a massy acoustic
fluctuation identified with the inflaton.  The energy that drives the
inflationary expansion is essentially that of the inertia of the inflaton
($=1/2 \mu^2\phi^2)$  which decays from $\phi=O(1)$ to the adiabatic
era. This mean energy within the patch is accompanied by small rapid
fluctuations. How these might be related to CMBR observed fluctuations is
an interesting question which remains to be elucidated in terms of the
causet model.

As mentioned, the advantage of the inflaton scenario is that inflation
stops naturally and passes over to adiabatic expansion. As this happens,
the exchange between the causets and modes continues and the mechanism of
Ref.1 to explain $\rho_\Lambda$ takes over.

A problem that arises in Ref.1 is that in the adiabatic epoch there exist
runs in which $\rho_\mathrm{Tot}$ goes negative since the fluctuations are
about mean situations for which $\rho_\mathrm{Tot}$ is close to zero,
$\rho_M$ being so small. If the above concepts have any validity, when
this happens the exchange equilibrium breaks down and the causets will
pump out more than they absorb, keeping $\rho_\mathrm{Tot} \geq 0$ and
$H^2 \geq 0$. Presumably such runs do not represent the cosmic environment
in which we live. But the main point to stress is that their existence
does not seem to pose a conceptual problem. They simply give rise to
situations too far from equilibrium to permit the applicability of the
pure random exchange process.

A far deeper problem is the target state with $\rho_\Lambda\to 0$ as
$\rho_M\to 0$, the quiescent metastable vacuum.  This is the problem
of the existence of a (metastable) vacuum with zero cosmological
constant, a highly controversial issue.

In terms of the above concepts, there is little one can say to
justify this ansatz save, as we have mentioned, its consistency
which bears repetition. One goes to the limit $\rho_M\to 0$, then it
must be supposed that $H\to
 0$, for if $H \neq 0$  the temporal variations of the metric will induce
the excitation of modes to make quanta, hence $\rho_M \neq 0$. But if
$H= 0$ then $\rho_\Lambda = 0$ since $H^2 = \rho_\mathrm{Tot}$.

$\rho_\Lambda=0$ means that the vacuum energy in the mode sector must
vanish on the average. This can be rationalized in that the conventional
zero point energy of modes must be combined with the gravitational energy
of the modes, both their mutual interactions and their interactions with
causet elements. It is easy to make models to shown how in, say a
Hartree-Fock approximation, one can choose a cut-off to make
$\rho_\Lambda=0$. But the question is, why is this cut-off so chosen, or
in other words, why does the cut-off take on a value which yields a
universe at absolute rest? There is something missing in our formulation
of physics which would make this ansatz compelling.

Of course this target space is metastable and from it new universes
can develop from fluctuating seeds. We exist within one of those.

\Section{Black Hole Evaporation}

Hawking's seminal paper on black hole evaporation was based on free field
theory, Ref.\cite{8}. The modes of a quantum field propagating in a
background Schwarzschild metric manifest themselves as conversion from
vacuum fluctuations of the past to on-mass-shell quanta in the future.

One of the principal problems that one encounters is that the field
configurations which give rise to this radiation adhere
exponentially closely to the horizon of the black hole all the way
into the past before they break away to become quanta with some
finite probability. Concomitantly their proper energy is
exponentially large, i.e. one is dealing with field configurations
in vacuum whose momenta are $>>1$. This is the transplanckian problem.
Clearly the idealization for using modes and free field theory
breaks down.

This does not mean that Hawking evaporation does not take place. Its
origin can be argued almost on thermodynamic grounds. If the matter that
makes up the black hole is placed within a cavity then one is dealing with
an eternal black hole. Green's functions of fields far from the horizon
are periodic in imaginary time as if there was a temperature there. And
this temperature is the same as that obtained using Hawking's free field
scenario for the production of radiation. See, for example, Ref.\cite{9}.

The use of the eternal black hole and the induced temperature does
not rely on free field theory, but only on very general properties
of quantum field theory. It also implies radiation since one can
imagine punching some holes in the boundary of the cavity to let
some radiation escape. It escapes at the Hawking temperature. It
{\it is} Hawking radiation and free field theory is not
necessary to get it.

Therefore we must inquire: how does one get around the transplanckian
problem in the Hawking process, the process that is applicable to the
situation of the dynamical collapse resulting in black hole formation? One
way is suggested by the causet scenario. Within a distance of $O(1)$ from
the horizon, space-time exists as causet elements and field configurations
cannot be described in terms of modes.  Successively, the field
configurations locked into the causets pick up the cisplanckian components
which get converted into modes. This happens when the wave functions of
the fields begin to overlap with the grainy structure of the space that
had been lain down in the prior history of the collapsing matter that made
the black hole. Then the Hawking mechanism takes over.

Parentani, Ref.\cite{10}, has shown how in the backward extrapolation of a
wave packet that is constructed to be an evaporated quantum the
interaction of that packet with incoming vacuum fluctuations causes the
packet to dissipate in the backward direction. We presume that it is in
the space-time region of this dissipation that the field configuration has
emerged from its sequestration within a causet element.

Parentani's demonstration shows that this rate of emergence into
modes and subsequent conversion to quanta is a steady state process
whose rate is that calculated by Hawking. So in the causet picture
there is a steady conversion of field from their configurations in
causet elements into modes.

At the same time the volume of space-time around the horizon increases due
to the reduction of the black hole mass i.e. that portion of space-time
that is within the horizon decreases according to Hawking's result $dM/dt
=M^{-2}$. This increase of the exterior Schwarzschild space may be
attributed to the increase in the number of causet elements which have
exchanged their field energy with the modes that have been converted to
quanta. Once more, as in inflation, a wake of causets is suggested by the
dynamics of energy creation. The slow roll of inflation is replaced by the
slow decrease of $M$.
$$$$

It is a pleasure and an honor for me to have had the occasion to
contribute these speculations, which were stimulated by the causet
mechanism of dark energy generation, to the $60^\mathrm{th}$ birthday
anniversary volume dedicated to Rafael Sorkin. This is to express my
gratitude to Rafael for his patient explanations and moreover to express
my admiration for this highly original and astute scientist. These pages
were written during my stay at the Perimeter Institute to which I express
my gratitude for its support.

\pagebreak


\begin{thebibliography}{}

\bibitem{1}M. Ahmed, S. Dodelson, P. B. Greene and R. Sorkin. Phys. Rev. {\bf D69}, 103523 (2004)
\bibitem{2}W. Kolb and M. S. Turner, {\it The Early Universe}, Perseus 1993 (Chapter 8)
\bibitem{3}A. Casher and F. Englert. Phys. Lett. {\bf 104B}, 115 (1981)
\bibitem{4}A. Linde, {\it Particle Physics and Inflationary Cosmology},  Harcourt Academic Publishers, Chur (1990)
\bibitem{5}R. Brout, F. Englert and E. Gunzig. Gen. Rel. Grav. {\bf 10}, 1 (1979) (1st prize Gravity Award Essay  (1979),
       A. A. Starobinski. Phys. Lett. {\bf 91B}, 99 (1980)
\bibitem{6} See the slow roll mechanism
      A. Albrecht and P. Steinhardt. Phys. Rev. Lett. {\bf 48}, 1220 (1982)
      A. Linde Phys. Lett. {\bf 108B}, 389 (1982)

\bibitem{broutx} R. Brout, August 2005 eprint gr-qc/0508019
\bibitem{8}S. W. Hawking, Comm. Math. Phys. {\bf 43}, 199 (1975)
\bibitem{9}R. Brout, S. Massar, R. Parentaini, Ph. Spindel., Phys. Rept. {\bf 260} (1995)
\bibitem{10}R. Parentani, unpublished 2005

\end{thebibliography}
\end{document}